\documentclass[twocolumn,
showkeys,preprintnumbers,amsmath,amssymb,prl]{revtex4}
\usepackage{graphicx}
\usepackage{dcolumn}
\usepackage{bm}
\usepackage{epsfig}
\usepackage{float}
\usepackage{amsmath,mathtools}
\usepackage{setspace}

\usepackage{tensind}
\tensordelimiter{?}

\usepackage{setspace}
\usepackage{color}
\usepackage{mathrsfs}

\begin{document}

\newcommand{\bs}{\boldsymbol}
\newcommand\eqn[1]{\begin{eqnarray} #1 \end{eqnarray}}
\newcommand\vect[1]{\boldsymbol{#1}}
\newcommand\mat[1]{\mathsf{#1}}
\newcommand\trans{^\mathsf{T}}

\newcommand\braket[3]{\left<#1\,\right|#2\left|\,#3\right>}
\newcommand\ket[1]{\left|\,#1\right>}
\newcommand\bra[1]{\left<#1\,\right|}
\newcommand\braketsc[2]{\left<#1\,|\,#2\right>}
\newcommand\eg{\emph{e.g. }}
\newcommand\ie{\emph{i.e. }}
\newcommand\nn{\mathrm{N}_2}
\newcommand\hh{\mathrm{H}_2}
\newcommand\methane{\mathrm{CH}_4}
\newcommand\ris{\{\vect{r}_i\}}
\newcommand\rris{\{\vect{R}_i\}}
\newcommand\dris{d\vect{r}_i^N}
\newcommand\x{\vect{x}}
\newcommand\xis{\{\vect{x}_i\}}
\newcommand\xxis{\{\vect{X}_i\}}
\newcommand\dxis{d\vect{x}_i^N}

\title{On the self-interference in electron scattering: Copenhagen, Bohmian and geometrical interpretations of quantum mechanics}
\author{Ivano Tavernelli} 
\email{ita@zurich.ibm.com}
 \affiliation{IBM Research -- Zurich, 8803 R\"uschlikon, Switzerland}
\date{\today}

\begin{abstract}
Self-interference embodies the essence of the particle-wave formulation of quantum mechanics (QM). 
According to the Copenhagen interpretation of QM, self-interference by a double-slit
requires a large transverse coherence of the incident wavepacket such that it covers the separation between the slits.
Bohmian dynamics provides a first step in the separation of the particle-wave character of matter by introducing
deterministic trajectories guided by a pilot wave that follows the time-dependent Schr\"odinger equation.
In this work, I present a new description of self-interference using the geometrical formulation of QM introduced in \textit{Annals of Physics} \textbf{371}, 239 (2016).
In particular, this formalism removes the need for the concept of wavefunction collapse in the interpretation of the act of measurement i.e., the emergence of the classical world.
The three QM formulations (Schr\"odinger, Bohmian, and geometrical) are applied to the description of the scattering
of a free electron by a hydrogen atom and a double-slit. 
The corresponding interpretations of self-interference are compared and discussed.
 \end{abstract}


\maketitle

\section{Introduction}
Quantum dynamics describes the time evolution of quantum objects that occurs at the microscopic scale, even though macroscopic
quantum effects are also well known, e.g., in superconductivity, superfluidity, lasers and circuit quantum electrodynamics.
The characteristic properties of quantum mechanics (QM) that have no classical counterpart 
can be traced to state superposition, quantum coherence and entanglement in composite systems.
All these phenomena have a natural explanation within the wavefunction representation
of Schr\"odinger quantum mechanics. 
However, when dealing with quantum measurements and the emergence of the classical
world, the theory needs to be complemented with an interpretative picture.
Among the most widely accepted ones is the Copenhagen interpretation of quantum measurement
according to which any quantum system has no well-defined properties prior to being measured.
It is during the measurement process that the wavefunction collapse occurs, producing a
deterministic outcome from a quantum-mechanical probability distribution.
Several alternative interpretations of quantum mechanics were proposed in the past, including
 the de Broglie--Bohm pilot-wave approach~\cite{deBroglie25,deBroglie26b,Bohm1952,Bohm1952a}, 
the many-worlds theory~\cite{Everett1957}, and the 
 geometrical interpretations~\cite{tavernelli_geom}.
Particularly relevant for this study are the trajectory-based formulations of quantum
 mechanics based on de Broglie--Bohm and quantum geometrization theories, which describe
 quantum processes by means of deterministic trajectories guided by a pilot wave or
 space curvature, respectively.

Since the very beginning of the quantum revolution at the turn of the last century, several 
fundamental  \textit{physical} as well as \textit{gedanken} experiments were designed to 
shed light onto the nature of purely quantum phenomena, such as state superposition (Schr\"odinger's cat), 
wavepacket coherence and (self) interference (e.g., in the Young's double-slit experiment).
The self-interference of a scattered electron on a double-slit described  by Feynman~\cite{Feyn_LN}
was realized experimentally by 
Davisson and Germer in 1927~\cite{Davidsson1928}, 
Merli, Missiroli and Pozzi in 1974~\cite{Merli1976},
and Tonomura and co-workers in 1989~\cite{tonomura1989demonstration}.
In these experiments, electrons are shot one at a time through the analogous of a double-slit 
and, even though they cannot interact with each other, the collection of the signals generated 
over time by the single electrons produce the typical interference pattern of wave mechanics.
Responsible for the self-interference is the wave nature of matter that is at the basis of the particle-wave
description of quantum mechanics within the Copenhagen interpretation.
However, the same experiment can also be explained within the Bohmian and the geometric formulations of quantum 
mechanics. 
In these cases, the wave character of the particles dynamics is described by a pilot wave or a space curvature
that guide the single trajectories to form the interference pattern at the detector plate.

Recently, the same self-interference experiment was repeated 
with 
atoms~\cite{Carnal1991}, 
large molecular systems,
such as carbon buckyballs (C60)~\cite{Zeilinger1999},
and large organic molecules (phthalocyanine) scattered on a nanograting with a spatial period of 10--100 nm~\cite{juffmann2012real}.
In all these examples, the molecular de Broglie wavelength ranges between 2 to 5 pm and 
is therefore several orders of magnitude smaller than the grid spacing.
In addition, the actual classical molecular diameter (which is on the order of 1 nm) is much larger than the
particle de Broglie wavelength.
These facts pose interesting challenges to the theoretical interpretation of the single-molecule self-interference 
process and to the validity of the particle-wave duality for massive particles.

The following basic quantities are relevant to all these investigations:
(\textit{i}) The de Broglie wavelength, $\lambda_{dB}$, which is defined as $h/p$, where $h$ is Planck's constant and $p$
is the modulus of the system's momentum. 
In the case of a polyatomic molecular system, the de Broglie wavelength is defined by the system's total mass and its center of mass velocity.
(\textit{ii}) The physical dimension of the slits (grid), characterized by the number of slits ($n_s$), their spacing ($d_s$) and
their width ($w_s$).
(\textit{iii}) The support of the molecular wavepacket  $\Psi(\bs r, \bs R,t)$ in space and time
(where $\bs r$ and $\bs R$ are the collective electron and nuclear coordinates, respectively),
which is defined (at each time $t$) as the region in space characterized by a nuclear density larger than
a given threshold $\epsilon$.
(\textit{iv}) The so-called transverse coherence of the molecular wavepacket $l_c=\lambda_{dB} L /\Delta x$, where
  $L$ is the distance between the source and the plane of the slits and $\Delta x$ is the size of the source~\cite{Juffmann2013}.
In the self-interference experiment on molecules, the \textit{static} molecular wavefunction   
will have a support with a diameter that is orders of magnitudes smaller than the grid spacing $d_s$,
and therefore -- independently of the size of the associated de Broglie wavelength -- 
it would be impossible to explain the self-interference process. 
In fact, double-slit interference patterns are described by the simultaneous emission of waves from `virtual' 
sources placed at the center of the slits. 
However, this can only occur if the incoming wavefront spans the transversal separation  of the two slits.
Within the `conventional' interpretation of QM, this is possible when the wavepacket is emitted from a narrow source,
which according to the Heisenberg relation $\delta x \delta p \geq \hbar/2$, will induce an uncertainty in the transverse 
particle momentum, leading to a macroscopic transverse wavepacket coherence of the size of the slit spacing.


This article is organized as follows. 
After an initial discussion on the different formulations of quantum dynamics and its interpretations, in the Theory section I briefly summarize the geometrical interpretation introduced in reference~\cite{tavernelli_geom}, putting particular emphasis on the correct definition of the curvature tensor in Finsfer spaces.
In the Method section, I introduce the equations of motion for the propagation of the electronic wavefunctions within the 
framework of time-dependent density functional theory (TDDFT) and discuss the derivation of the trajectory-based quantum approaches (Bohmian and geometric).
These methods will then be used to study the scattering of a single electron 
with a hydrogen atom and with a double-slit. 
The aims of this study are 
(\textit{i}) 
to show how the geometrical approach introduced in~\cite{tavernelli_geom} is compatible with the known theoretical and experimental results and 
(\textit{ii}) 
to illustrate the emergence of self-interference pattern within these fully deterministic frameworks.
Finally, I will conclude with a comparison between the different formulations of self-interference:
wavefunction-based (Copenhagen interpretation), Bohmian (trajectory-based) and geometrical (geodesics-based).

\section{Deterministic formulations of quantum dynamics}
In this paper, I present a theoretical study of the scattering dynamics of an electron with a
standing hydrogen atom and a double-slit. 
The results are analyzed using the two  deterministic approaches introduced in the previous section.
(\textit{i}) Bohmian dynamics:
within this picture, electrons are described by deterministic trajectories driven by the pilot wave.
Starting from the electronic wavefunction, 
one can derive the Bohmian paths~\cite{curchod2013ontrajectory,Curchod2011,tavernelli2013ab}, which are
perfectly consistent with the wavefunction representation of quantum dynamics~\cite{hollandbook,wyatt2002,Duerr2009,sanzbook2012,Benseny14}. 
Note that despite the different ontological content of the Schrodinger and Bohmian 
representations of quantum dynamics, their physical content is equivalent.
In fact, quantum trajectories can be derived without the need of further assumptions directly and uniquely 
from the solution of the time-dependent Schr\"odinger equation (TDSE), whereas the density of the quantum trajectories in configuration space provides
an exact description of the quantum density of the system (which according to the Hohenberg--Kohn theorem~\cite{hohenberg64},  
one can derive all observables).
(\textit{ii}) The geometrical approach introduced in Ref.~\cite{tavernelli_geom}:
within this framework, the quantum dynamics is described by means of trajectories moving in a curved configuration space,
where the metric of the space is a function of the positions and momenta of all particles in the system.
This is done in a Finsler manifold~\cite{Rund59}, which can be interpreted as an extension of a Riemann space. 
All quantum effects are therefore absorbed into the geometry of the space in a similar way as
in general relativity, where masses induce a curvature of space-time and particles evolve along geodesics. 
As in the case of Bohmian dynamics, this formulation of quantum mechanics contains all information
inherent in the Schr\"odinger wavefunction representation with, however, the addition of a further `variable',
namely, a  quantum trajectory that introduces determinism in the picture.
In addition, this geometrical interpretation of quantum dynamics provides a unique and unambiguous classical
limit~\cite{tavernelli_geom}. 


\section{Theory}

Within the geometrical interpretation of quantum dynamics introduced in~\cite{tavernelli_geom}, matter points evolve along \textit{single} quantum trajectories described by the geodesic of the curved phase 
space.
This theory is strictly deterministic, as it does not rely on the probabilistic interpretation of the 
quantum-mechanical wavefunction.
In fact, differently from Bohmian dynamics, the particle position is not described by a probability distribution,
but is fully localized, whereas the wavefunction nature of quantum theory is absorbed into the geometry of the 
metric space. 
The dynamics takes place in the extended configurations space that includes time $t$ 
(and therefore has dimension $2\,(3N+1)$), whereas the progress of the dynamics is measured in terms of a \textit{proper} time parameter, $\tau$.
In this article, I will use the symbol $x=(x^0,\dots,x^N)$ for the collective variables $(x^0=t,\bs q_1,\dots, \bs q_N)\equiv (x^0,q)$  and $y=(y^0,\dots,y^N)$ for
$(\dot x^0, \dot{\bs q}_1,\dots, \dot{\bs q}_N) \equiv (y^0, \dot{q})$  with $y^0=\partial t/ \partial \tau$.
For any given initial condition, the quantum dynamics associated with each configuration point follows a 
deterministic trajectory in the curved $3N+1$-dimensional space according to the geodesic curve $\tau \mapsto \gamma(\tau)$
(using Einstein's summation convention and $a, b,  c=0,\dots, 3N$)
\begin{equation} \label{prop1} 
\ddot{\gamma}^{a}+?\Gamma^{a}^{}_{b  c}?(\gamma,\dot\gamma) \dot{\gamma}^{b} \dot{\gamma}^{ c} = - {g}^{a b} \partial V(q) /\partial \gamma_b \, ,
\end{equation}
where 
$?\Gamma^{a}^{}_{b  c}? =\frac{1}{2} {g}^{ad}({g}_{d  c,b}+{g}_{d b, c}-{g}_{ b  c,d})$ are the generalized connections 
(with ${g}_{a b, c}=\partial_{c}{g}_{ab}$ and $\dot\gamma=\partial_{\tau}\gamma(\tau)$), 
${g}_{a b}$ are the metric coefficients 
\begin{equation} \label{prop2}
g_{a  b}(x,y)=\frac{1}{2} \frac{\partial^2 \Lambda^2(x,y)}{\partial y^{a} \partial y^{ b}} \, ,
\end{equation}
with 
\begin{equation} \label{prop2b}
\Lambda(x,y)=\mathcal{T} (\dot{q})/y^0  - Q(q) y^0 \, 
\end{equation}
and $\mathcal{T} (\dot{q})=(1/2)  \, m_e \, \sum_i^N\dot{ \bs q}_i^2$ ($m_e$ is the electron mass).
In Eqs.~\eqref{prop1}-\eqref{prop2b}, I use $\tau=s$, where $s$ is the arclength defined by $ds=\Lambda(x,y) \, d\tau$; 
$x(t)$ and $y(t)$ are explicit functions of time, $V(q)$ is the classical potential, and $ Q(q)$
is the quantum potential. 
The conditions that the extended Finsler function $\Lambda(x,y)$  needs to fulfil 
are discussed in Refs.~\cite{tavernelli_geom,Rund59}. These are:
(\textit{i}) positive homogeneity of degree one in the second argument, $\Lambda(x,k y)= k \Lambda(x, y), k>0 $, 
(\textit{ii}) $\Lambda(x,y) > 0$ with $\sum_i (y^i)^2 \neq0$, and 
(\textit{iii}) $\frac{1}{2} \frac{\partial^2 \Lambda^2(x,y)}{\partial y^{a} \partial y^{b}} \xi^{a} \xi^{b} > 0, \forall {\xi}\neq \lambda y$. 
In Appendix A, I will discuss the validity of these conditions for $\Lambda(x,y)$ defined in Eq.~\eqref{prop2b}.

In Ref.~\cite{tavernelli_geom}, I defined the curvature in the Finsler space manifold as 
\begin{equation}
R(x,y)= g^{a b}(x,y) R_{a b} (x,y) \, ,
\end{equation}
where $R_{a b}=?R^{c}^{}_{a b c}?$ is the Ricci tensor and 
 \begin{align} \label{Eq_riemann_curv}
?R^{d}^{}_{a b c}?(x,y) =&
?\Gamma^{d}^{}_{a c, b}? (x,y) -?\Gamma^{d}^{}_{a b, c}? (x,y)+ \notag \\
& ?\Gamma^{d}^{}_{b s}? (x,y) ?\Gamma^{s}^{}_{ac}? (x,y)-
?\Gamma^{d}^{}_{c s}? (x,y) ?\Gamma^{s}^{}_{ab}? (x,y)
\end{align}
is the Riemann curvature tensor with $?\Gamma^{d}^{}_{a c, b}? = \partial ?\Gamma^{d}^{}_{a c}?/\partial x^{b}$. 
However, this is only the case when the Finsler space (M, $\Lambda$) is a metric space with $g_{a b}(x)$
a function of the coordinates $x$ of the base manifold M (i.e., the extended configuration space).
A detailed account on the derivation of the Cartan non-linear curvature in general (non-metric) Finsler spaces and the related linear connection is given in Appendix B and references~\cite{Rund59,Thesis_Pfeifer}. 
Here, I just introduce the definition of the linear Cartan curvature tensor~\cite{Thesis_Pfeifer} (removing the dependence on $x$ and $y$)
\begin{align}
^l\hspace{-0.07cm}?R^d_{}{abc}?= & \delta_b ?{{\tilde\Gamma}}^d_{}{ac}? - 
\delta_c ?{{\tilde\Gamma}}^d_{}{ab}? + 
?{{\tilde\Gamma}}^d_{}{sb}? ?{{\tilde\Gamma}}^s_{}{ac}? - 
?{{\tilde\Gamma}}^d_{}{sc}?  ?{{\tilde\Gamma}}^s_{}{ab}? - \notag \\
& ?C^q_{}{as}? ?R^s_{}{bc}?.
\end{align}
where
$
?{{\tilde\Gamma}}^c_{}{ab}? = \frac{1}{2} g^{cq} (\delta_a g_{bq} + \delta_b g_{aq} - \delta_q g_{ab})
$ 
are the coefficients of the Cartan linear covariant derivative, 
$\partial_{a}$ are the partial derivatives with respect to the coordinates,
$\delta_{a}=\partial_{a}-?N^{b}^{}_{a}? \bar{\partial}_{b}$ are their horizontal components, 
and $\bar\partial_{a}$ are the corresponding derivatives in the vertical tangent space (partial derivatives in the velocities). 
$?N^{b}^{}_{a}?$ are the Cartan connection coefficients,  
$
C_{abc}=\frac{1}{4} \bar\partial_a \bar\partial_b \bar\partial_c \Lambda^2  \, ,
$
and
$ 
?R^a_{}bc?=\delta_c ?N^{a}_{}b? - \delta_b ?N^{a}_{}c?
$
is the Cartan non-linear curvature.
The corresponding geodesic equation becomes
\begin{equation}
\ddot{\gamma}^a + ?N^{a}_{}b?(\gamma,\dot{\gamma}) \dot\gamma^b= 
- {g}^{a b} \partial V(r) /\partial \gamma_b \, ,
\end{equation}
which is equivalent to the one of Eq.~\eqref{prop1} (see Appendix B).

In addition to the geodesic equation for the propagation of the trajectory (Eq.~\eqref{prop1}), a second  
equation of motion for the time evolution of the space curvature is required. 
This is achieved by defining an energy matter tensor whose nature and dynamics are discussed
in~\cite{tavernelli_geom,hollandbook}.
For the practical purposes of this work, the following statement (proven in~\cite{tavernelli_geom}) is of particular relevance:
For a given initial point in phase space, $( q(0), \dot{ q}(0))$, the geodesic dynamics described by Eq.~\eqref{prop1} is
 equivalent to the Bohmian trajectory defined by 
 the velocity field 
 \begin{equation}
\bs v^{\Psi}_k(x)=\frac{\hbar}{m_k} \Im \left[ \frac{\nabla_k \Psi(x)}{\Psi(x)} \right]
 \label{eq:vel_field}
 \end{equation}
 and started from the same point. In Eq.~\eqref{eq:vel_field}, $\Psi(x)$ is the system wavefunction and $m_k$ is the particle mass.
 
As in the Bohmian case, also in this formulation there is a unique trajectory unambiguously assigned to the system dynamics.
In particular, each particle in the system contributes to the total curvature of the configuration space. 
For a single isolated particle with zero momentum the curvature does not trigger any dynamics; 
however, when other particles are present, the overall curvature induces the dynamics of the different particles constituting the system, which can be decomposed into a set of individual, but correlated, trajectories in the Euclidean space.

\section{Methods}
\begin{figure}[h]
\begin{center}
\includegraphics[width=8 cm]{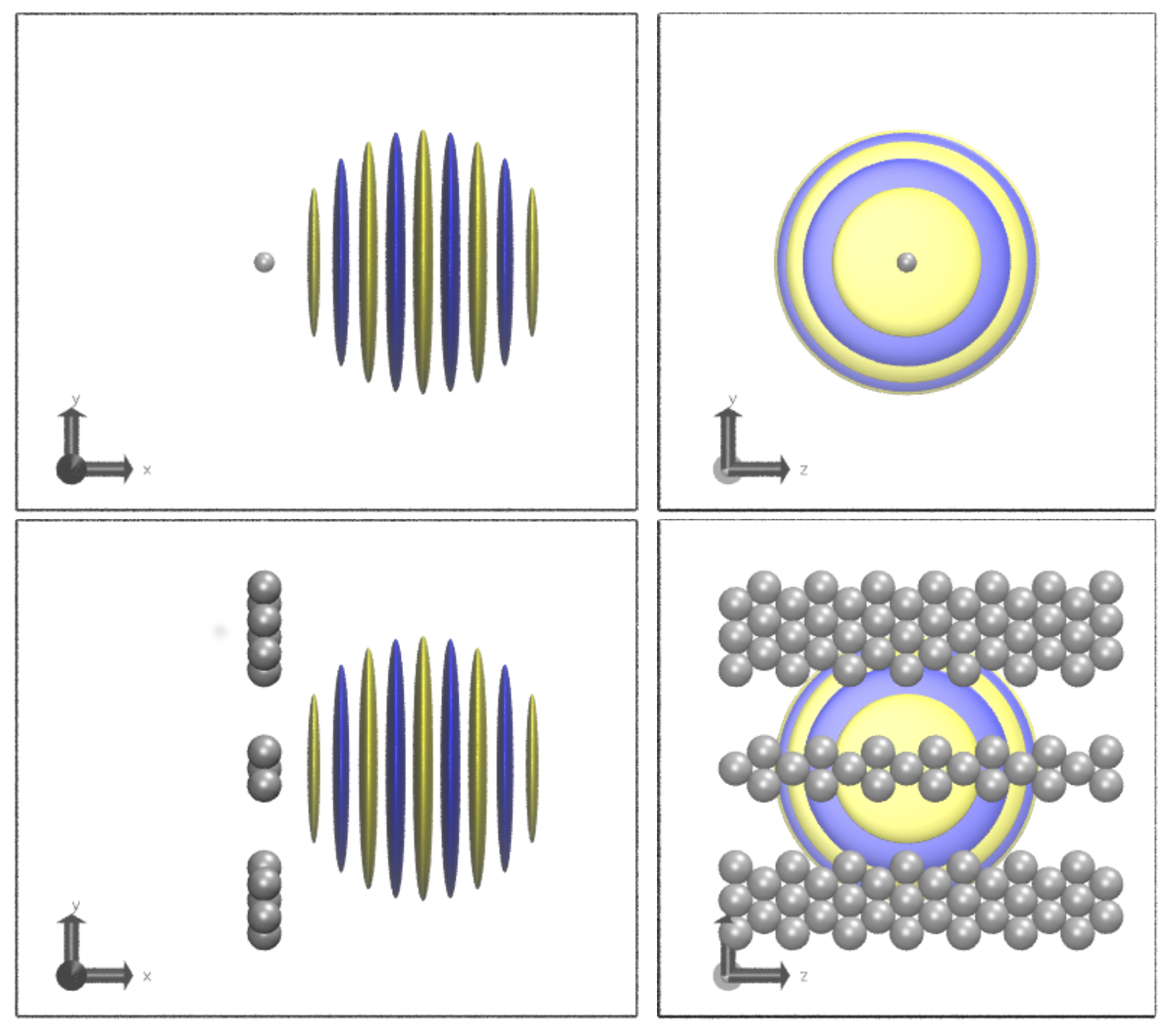}
\caption{Side and front views of the initial states of the dynamics. Upper panels: 
Scattering of an electron by a hydrogen atom. 
Lower panels:  Scattering of an electron by double-slit obtained from a zinc surface. 
The pictures show the real part of the initial Gaussian
wavepacket in the $xy$- and $yz$-planes. The electron is initiated with a velocity
of $2.18 \, 10^6$ m/s \ in the $-x$ direction.}
\label{figure1}
\end{center}
\end{figure}

\paragraph{System preparation.}
In this study, I consider the scattering of a free electron with a standing hydrogen atom and a double-slit, each of width $ w_s\simeq 8$~\AA~ and separated by a distance $d_s\simeq 3$~\AA, carved in a monolayer slab of zinc atoms (see Fig.~\ref{figure1}). 
All nuclei are treated classically, whereas the electrons (bound and free) are described quantum mechanically using DFT.
Zinc atoms were chosen because of the possibility to obtain a high electronic density using just two valence electrons per atom and a pseudopotential~\cite{goedecker96} for all other `core' electrons. However, the results do not depend on the nature of the 
slab composition.

The free electron is described by a Gaussian wavepacket (see Fig.~\ref{figure1})
\begin{equation}\label{eq:phiG}
\phi_G(\bs{q}_1,\bs{k}_0)=\left(\frac{\sigma^2}{2\pi^3}\right)^{3/4} \int d^3p \, e^{-\sigma^2 (\bs{k}-\bs{k}_0)^2} e^{-i \bs{k} (\bs{q}_1-\bs{q}^0_1)}
\end{equation}
with $\sigma=2.65$ \AA \ and an initial wave vector of length $k_0=2.83$ \AA$^{-1}$ pointing in the direction of the hydrogen atom. 
The electronic wavepacket has therefore a de Broglie wavelength of $2.22$ \AA.

In the case of the scattering on the hydrogen atom, the total electronic wavefunction in the singlet state can be 
separated into spatial and spin parts 
 \begin{align} \label{eq:singlet}
\psi(\bs{q}_1,\bs{q}_2; \bs{p}_0)=&\frac{1}{2\sqrt{1+S^2}} ( \phi_G(\bs{q}_1,\bs{p}_0/\hbar)\phi_{1s}(\bs{q}_2) +\notag \\& \phi_G(\bs{q}_2,\bs{p}_0/\hbar)\phi_{1s} (\bs{q}_1)) \, 
 (\alpha_1 \beta_2 - \beta_1 \alpha_2) \, ,
\end{align}
where (at time $t=0$) $\phi_{1s}(\bs{q})$ is the hydrogen \textit{1s} orbital, $\bs{p}_0=\hbar \bs{k}_0$ and $S$ is the overlap integral between $\phi_G(\bs{q},\bs{p}_0/\hbar)$ and $\phi_{1s}(\bs{q}) $. 
When the Gaussian wavepacket is perfectly aligned with the hydrogen atom (with the electron momentum parallel to the electron-proton distance vector),  the impact parameter $b$ is zero.
Because of the symmetry of the system, each displaced Gaussian wavepacket (in the $yz$-plane of Fig.~\ref{figure1}) is characterized by a value 
of $b_G$ equal to the distance between the line associated to the velocity vector at the Gaussian center
and the position of the hydrogen atom.

In the case of the double-slit, the total wavefunction of the system (zinc surface and incident electron) cannot be described 
exactly within the DFT/TDDFT formalism. 
In this study, I consider the total Hilbert space to be the direct product between the Hilbert space of the electrons of the 
surface, $\mathcal{H}_s$, and the Hilbert space of the single scattered electron, 
$\mathcal{H}_e$, i.e., $\mathcal{H}_\text{tot}=\mathcal{H}_s \otimes \mathcal{H}_e$. 
Because of the high energy of the incident electron, I expect this approximation to hold along the entire trajectory, as for  the scattering with the hydrogen atom.
In both cases, all quantities involved in the propagation of the quantum trajectories, namely, the quantum potential and the
metric tensor, are evaluated from the time-propagated wavefunction of the scattered electron. 
However, it is worth mentioning that the time evolution of all orbitals in the systems (surface and incoming electron) feel 
the potential generated by the entire system, as described by the DFT/TDDFT representation of the many-body interactions.

\paragraph{TDDFT dynamics.}
Applying Dirac-Frenkel's variational principle~\cite{tavernelli06,curchod2013trajectory} 
to the TDDFT action functional subject to the constraint $\rho(\bs q,t)=\sum_{k=1}^{N_e} |\phi_k(\bs q,t)|^2$, one obtains the time-dependent Kohn--Sham equations (TDKS) or a system of $N_e$ electrons
\begin{equation}
i \hbar \frac{\partial}{\partial t} \phi_k(\bs q,t) = -\frac{\hbar^2}{2 m_e} \nabla^2 \phi_k(\bs q,t) +v_\text{s}[\rho, \Phi_0] (\bs q, t) \, \phi_k(\bs q,t) \, , 
\label{eq:tdks}
\end{equation}
where
\begin{equation}
v_\text{s}[\rho,\Phi_0](\bs q, t)=v_{\text{ext}}(\bs q) + v_H[\rho, \Phi_0] (\bs q, t) + \frac{\delta \mathcal{A}_{xc}[\rho,\Phi_0](\bs q, t)}{\delta \rho(\bs q,t)} \, ,
\label{eq:vs}
\end{equation}
$v_{\text{ext}}(\bs q)=-\frac{ e \, Z_H}{|\bs{R}_H-\bs{q}|}$ is the Coulomb potential of the proton, $e$ is the electron charge, 
$v_H[\rho, \Phi_0]$ is the Hartree potential and
$\Phi_0(\bs q,t_0)$ is the initial wavefunction at time $t_0$ with corresponding density $\rho(\bs q,t_0)$. To simplify the notation, in the following I remove the dependence on the initial value conditions.
In the so-called \textit{adiabatic approximation}, the functional derivative of the time-dependent exchange-correlation action functional in Eq.~\eqref{eq:vs} is approximated by
\begin{equation}
v_{xc}[\rho](\bs q,t)= \frac{\delta \mathcal{A}_{xc}[\rho]}{\delta \rho(\bs q,t)} \approx \left. \frac{\delta {E}_{xc}[\rho]}{\delta \rho(\bs q)}\right|_{\rho(\bs q)\leftarrow \rho(\bs q,t)} = v_{xc}[\rho](\bs q)
\label{eq:bc17.1}
\end{equation}
where $E_{\text{xc}}[\rho]$ is the DFT exchange and correlation energy functional.
The time-dependent KS equations in Eq.~\eqref{eq:tdks} are integrated using the unitary Cayley propagator~\cite{note_cayley,castro3425,curchod2013trajectory}. 
The numerical implementation of this propagation scheme in CPMD~\cite{cpmd} and 
its coupling to the (classical) nuclear dynamics  are described in full detail in~\cite{tavernelli2005molecular,Tavernelli2015}.

\paragraph{Bohmian trajectories.}
In Bohmian dynamics, the system wavefunction is interpreted as a pilot wave that guides the motion of the particles
in the configuration space. 
For a two-electron system (the generalization to many-electron systems is straightforward), the Bohmian velocity fields associated with their trajectories are
\begin{equation}
\begin{dcases} \label{theo1}
v_1(\bs{q}_1,\bar{\bs{q}}_2,t) =\frac{\hbar}{m_e} \Im \frac{\nabla_{\bs{q}_1} \psi(\bs{q}_1,\bar{\bs{q}}_2,t)}{\psi(\bs{q}_1,\bar{\bs{q}}_2,t)}\\
v_2 (\bar{\bs{q}}_1,\bs{q}_2,t) =\frac{\hbar}{m_e} \Im \frac{\nabla_{\bs{q}_2} \psi(\bar{\bs{q}}_1,\bs{q}_2,t)}{\psi(\bar{\bs{q}}_1,\bs{q}_2,t)} \, ,
\end{dcases} 
\end{equation}
\noindent where the bar in $\bar{\bs{q}}_i$ ($i=1,2$) indicates that for each numerical integration step this coordinate
is assumed constant. 
If the initial electronic probability is interpreted as the probability to find the unbound electron in a given
position $\bs{q}$ at $t=0$, then the swarm of Bohmian quantum trajectories describes the time evolution of a 
statistical ensemble of different possible initial coordinates~\cite{Sanz08}.
This implies that for each choice of the initial position of the center of the Gaussian wavepacket ($\bs{q}_1(0)$ 
in Eq.~\eqref{eq:phiG}),  there is always a finite probability to start a quantum trajectory with a impact 
parameter $b$ that differs from the displacement $b_G$ associated to the center of the Gaussian distribution. 
In Bohmian mechanics, the density of trajectory points at any time $t$ corresponds to the wavefunction probability
density obtained from $|\psi(\bs{q}_1,\bs{q}_2,t)|^2$. 
In this sense, Bohmian dynamics is completely consistent with the wavefunction formulation of quantum dynamics.
The generalization of this notation to the case of the scattering by a double-slit is trivial. 

\paragraph{The geometrical description.}
In the particular case of the two-electron system with coordinates $\bs q_1 \in \mathbb{R}^3$ and $\bs q_2 \in \mathbb{R}^3$, the dynamics takes place in the extended 
configuration space that includes time $t=x^{0} \in \mathbb{R}$, whereas the progress of the dynamics is measured in terms of a \textit{proper} time parameter, $\tau$.
For the description of the scattering of an electron by a hydrogen atom, I will use the symbol $x$ for the collective variables $(x^0,\bs q_1,\bs q_2) \in \mathbb{R}^7$ and $y$ for
$(y^0, \dot{\bs q}_1,\dot{\bs q}_2) \in \mathbb{R}^7$, with $\dot x^0=y^0=\partial t/ \partial \tau$.

\section{Results and discussion}

\paragraph*{The simplest possible realization of the self-interference experiment.}
The simplest possible self-interference (`double-slit') Tonomura's setup is obtained from the scattering of an electron
by a hydrogen atom (see Fig.~\ref{figure1}, upper panels).
For a small enough impact parameter, one obtains a splitting of the electron wavepacket into two
branches moving, respectively, to the right and to the left of the scattering atom, with the possibility to self-interfere later
on.
The total wavefunction in the singlet state is described by the product state in Eq.~\eqref{eq:singlet}.
Starting from the uncorrelated initial wavepackets  $\phi_{G}(\bs{q}_1, \bs{k}_0)$ and $\phi_{1s}(\bs{q}_2)$, 
I therefore propagate, at each time step, the one-electron Kohn-Sham orbitals according to Eq.~\eqref{eq:tdks}, and
then reconstruct the corresponding two-electron wavefunction using Eq.~\eqref{eq:singlet}.

\begin{figure}[h]
\begin{center}
\includegraphics[width=8.cm]{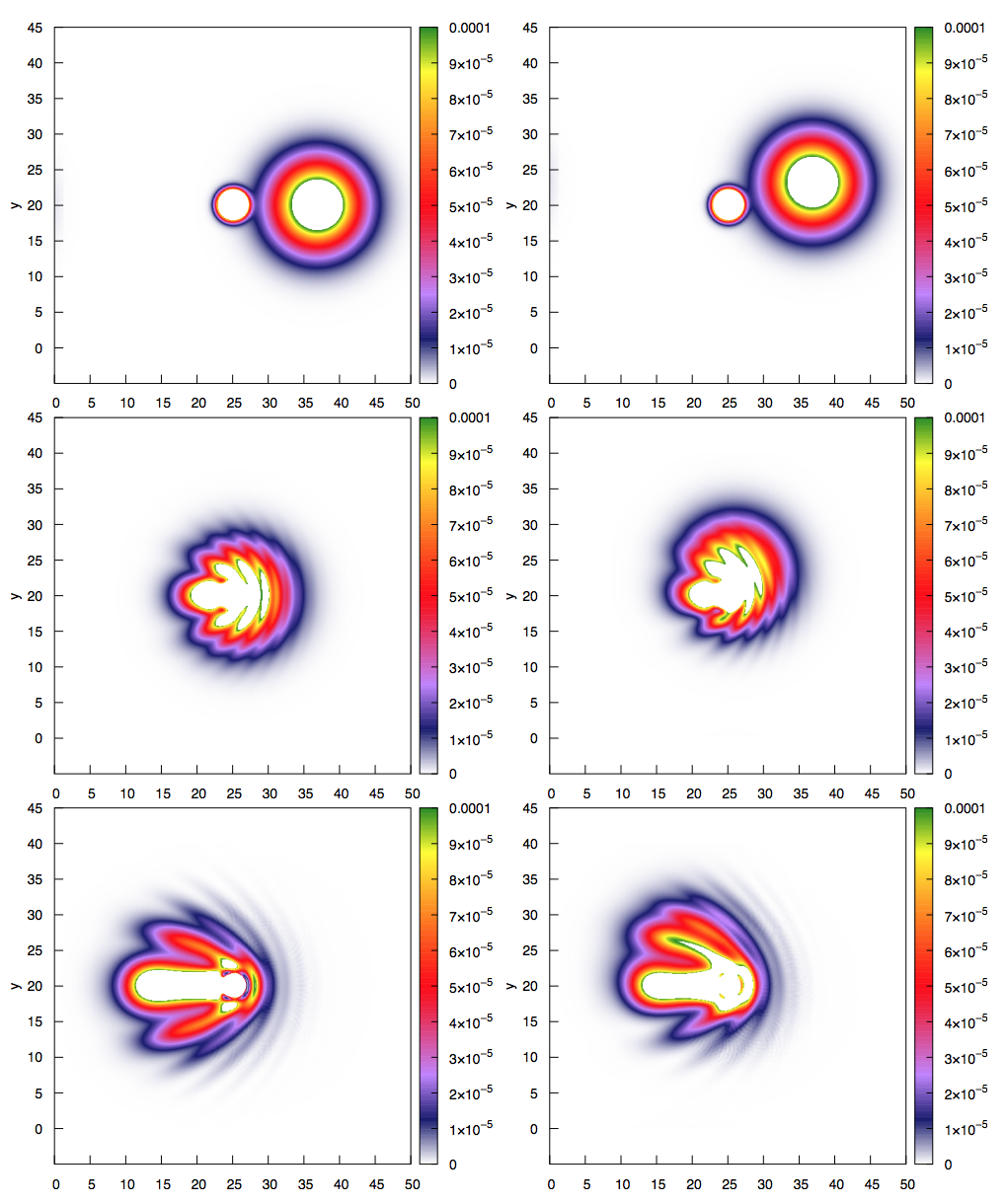}
\caption{Two-dimensional electron density profiles of the scattering of an electron by a hydrogen atom. 
The electron is directed towards the hydrogen atom placed at coordinates (25 \AA, 20 \AA, 20 \AA) with an initial velocity of $2.18\, 10^6$ m/s and an impact parameter $b=0$  \AA~ (left column) and  $b=3.1$  \AA~ (right column). From the top:  $t=0.048$, $0.363$ and $0.726$~fs.
All profiles are evaluated in the $xy$-plane that passes through the hydrogen atom. 
(Coordinate axes in \AA).
}
\label{figure2}
\end{center}
\end{figure}

Fig.~\ref{figure2} shows the slices of the total electronic density 
for a fixed position $\bs{q}_2$ at three different times:  $t=0.048$, $0.363$ and $0.726$~fs. 
The left column reports the results obtained with an impact parameter $b=0$ \AA, whereas on the right, the same density profiles are shown for $b=3.1$~\AA.
The incoming electron moves from the right to the left with an initial velocity of $2.18\, 10^6$ m/s, and before the collision takes place, it is described by a spherical-symmetric Gaussian distribution. 
Upon interaction with the hydrogen atom originally placed  at the center of the simulation box with zero velocity, the incoming
electron is scattered, producing a clear interference pattern that evolves into a series of minima and maxima of the electronic 
density. 
According to the Copenhagen interpretation of quantum mechanics, each scattering event will produce a single
spot at the detector plate at a position that is 
determined by the electronic probability density of the scattered electron. 
The cumulative pattern obtained by a sequence of independent scattering events will then reproduce the `self-interference'
picture measured by Tonomura et al.~\cite{tonomura1989demonstration} in their double-slit interferometer experiment.

In Bohmian dynamics, the electronic wavefunction is interpreted as a pilot wave that guides the dynamics
of the point particles~\cite{Bohm1952}. 
Formally, this theory is inherently a multiple-trajectory approach in the sense that the system wavefunction
also determines the probability distribution of the particle positions and momenta.
In numerical calculations, this property can lead to two different implementation of the dynamics:
in the first approach, the initial position distribution is used to sample the initial coordinates to start the dynamics,
whereas in the second approach, the trajectories are started from an arbitrary (usually regular) grid of points in configuration 
space and then propagated taking into account the amplitude associated to each trajectory.
In both approaches, the probability distribution obtained from the collection of all trajectory points at a later time 
(multiplied by the corresponding probabilities in the second approach) reproduces the square of the corresponding
quantum-mechanical wavefunction at the same time. 
The main advantage of the Bohmian representation of quantum dynamics is that it does not require the concept 
of wavefunction collapse for the interpretation of  measurements and in particular for the emergence of the classical 
limit. In other words, Bohmian dynamics is deterministic and therefore can naturally account for the observation
of single, localized scintillations in the scattering image of the Tonomura's experiment.
In this picture, self-interference is induced by the scattering of the electron wavefunction of the incoming electron
with the one of the standing hydrogen atom. 
Upon interaction, the scattered wavefunction develops a series of nodal lines that separate regions with high 
electronic amplitude (see Fig.~\ref{figure2}). 
Ultimately, the driving force that causes the trajectories to form the characteristic scattering pattern can be
associated with the action of the quantum potential. 

\begin{figure}[h]
\begin{center}
\includegraphics[width=8cm]{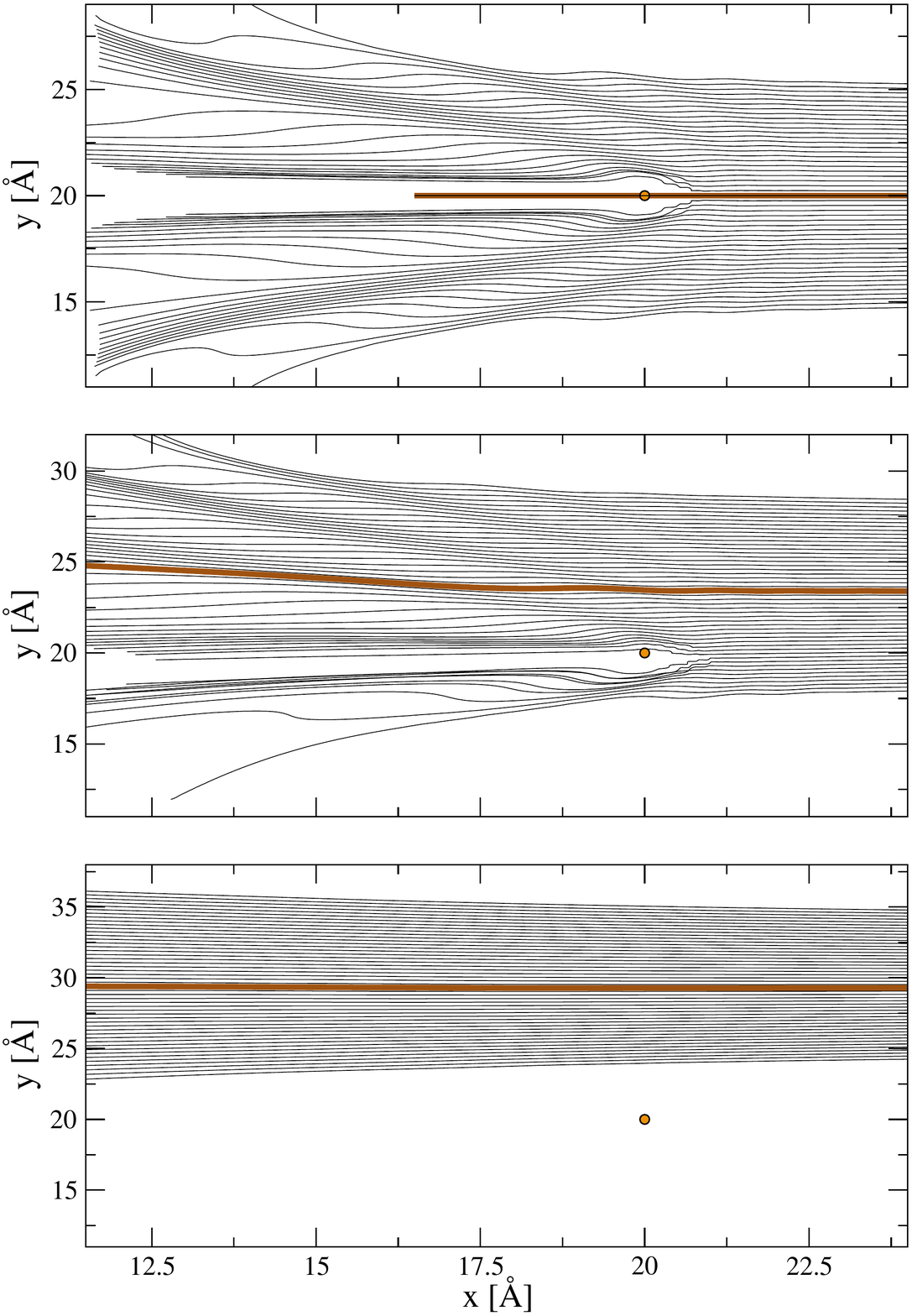}
\caption{Bohmian trajectories in the $xy$-plane obtained for the scattering of an electron by a hydrogen atom.
The trajectories are started from a regular 1D grid passing through the center of the Gaussian at $t=0$ and 
parallel to the $y$-axis. Top to bottom: $b=0, 3.5$ and $9$~\AA. The  circle indicates the position of the hydrogen atom.}
\label{figure4}
\end{center}
\end{figure}

In the Bohmian formulation of the scattering process, there is however an ambiguity concerning the 
interpretation of the impact parameter. 
In fact, for any choice of the impact parameter $b$, it is possible to associate different trajectories guided by
different wavefunctions. 
For example, one can consider a trajectory with an impact parameter $b_t$ guided by a Gaussian wavepacket 
characterized by  
an impact parameter of zero for its center, $b_G=0$.
In this case, the trajectory will correspond to a point in space with an associated amplitude smaller than the maximal value
at the center of the Gaussian.
However, one could also consider the dynamics of the central point of a Gaussian wavepacket for which  $b_t=b_G$. 
In this case, the trajectory is the one with the maximum possible amplitude.
Clearly, the dynamics of these trajectories follow different paths. 
In Fig.~\ref{figure4}, I show the dynamics of 50 trajectories started from a 1D regular grid oriented along the $y$-axis (see Fig.~\ref{figure1}) and passing through the center of the Gaussian wavepacket at $t=0$. 
The results are obtained for different values of the impact parameter $b_G$:  0 \AA~ (top panel), 3.5 \AA~ (middle panel), and 9 \AA~(bottom panel). 
Before scattering with the hydrogen atom, the trajectories proceed mainly in parallel, with a slight dispersion induced by the diffusion of the guiding wavepacket. 
After the collision, we observe a focusing of the trajectories to form regions of high density separated by regions of low density (nodal lines). 
Each trajectory describes the time evolution of a single electron started from a different position in space, which is compatible 
with the distribution of the original wavepacket.
Again, the same impact parameter, $b_t$, can be associated to the trajectory guided by a wavepacket initiated with $b_G=0$ (top panel) or to another trajectory driven by a Gaussian displaced from the origin, e.g. with $b_G=b_t$ (middle panel).
The signal produced by the ensemble of all trajectories (for a given value of $b_G$) on a plate placed in the $yz$-plane  perpendicular to the motion of the scattered electron will show the characteristic self-interference pattern of the Tonomura experiment. 
This example illustrates the strong interplay between the dynamics of the guiding wavepacket (Fig.~\ref{figure2}) and the one of the single-particle Bohmian trajectories (Fig.~\ref{figure4}).

\begin{figure}[h]
\begin{center}
\includegraphics[width=8.cm]{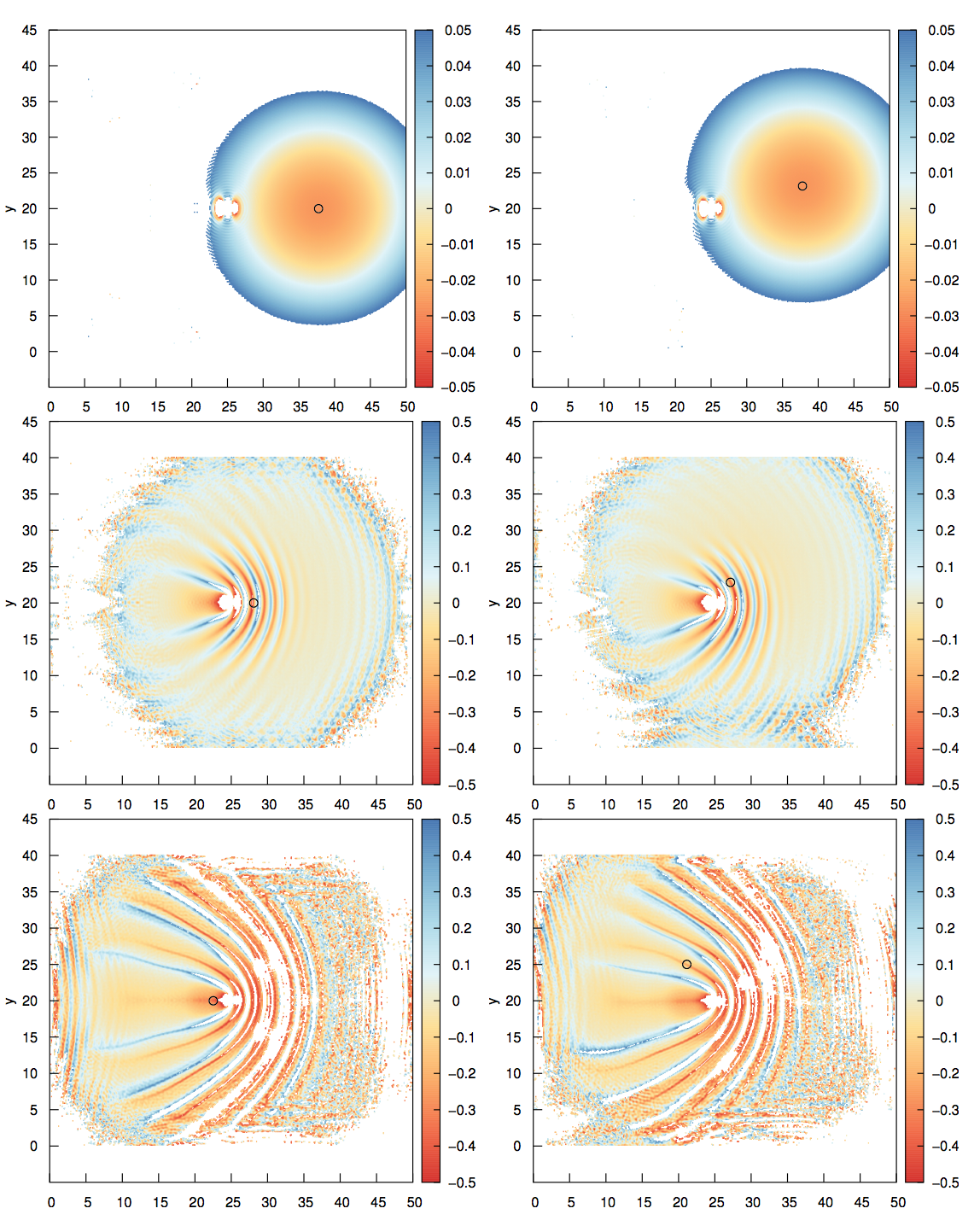}
\caption{Two-dimensional projections of the metric component $g_{11}(x,y)$ computed for the scattering of an electron by a hydrogen atom. The electron is directed towards the hydrogen atom placed at coordinates (25 \AA, 20 \AA, 20 \AA) with an initial velocity of $2.18 \, 10^6$ m/s and an impact parameter $b=0$~\AA~ (left column) and  $b=3.5$~\AA~ (right column). Top to bottom:  $t=0.048$, $0.363$ and $0.726$~fs. The black circles indicate the position of the electron in the different frames.
All profiles are evaluated in the $xy$-plane that passes through the hydrogen atom. (Coordinate axes in \AA).}
\label{Fig_qpot}
\end{center}
\end{figure}

According to the geometrical description of quantum dynamics given in Ref.~\cite{tavernelli_geom}, particles evolve along geodesic lines in the curved space, where the curvature is induced by the quantum potential.
Differently from the Bohmian dynamics case, there is a unique trajectory assigned to each particle unambiguously.
Fig.~\ref{Fig_qpot} shows the time evolution of the spatial component of the metric tensor $g_{11}(x,y)$ induced by the scattered electron, which is proportional to the quantum potential $Q(q)$ projected onto the $xy$-plane. 
(Also in this case, I use the notation $x= (x^0,q)= (x^0,\bs q_1, \bs q_2$) and $y= (y^0,\dot q)= (y^0,\dot{\bs q}_1, \dot{\bs q}_2$) 
and the fact that 
$
-\frac{1}{2} {\partial^2 \Lambda_{\mathcal{Q}}^2(x,y)}/{\partial y_i^2} = Q(q) \, m_e
$, where $\Lambda_{\mathcal{Q}}^2(x,y)= -2 \, T(\dot{q}) \, Q(q)$ is the dominant component of $\Lambda^2(x,y)$).
The left column shows  three different snapshots of $g_{11}(x,y)$ obtained at times 
$t=0.048$, $0.363$ and $0.726$~fs for a trajectory with an impact 
parameter $b_G=0$ and an initial velocity $v_G(0)=2.18 \, 10^6$ m/s., whereas the right column reports the same quantity for $b_G=3.5$~\AA.
At  $t=0.048$~fs, the scattered electron can be considered, in good approximation, an independent particle weakly interacting with the scattering center (the hydrogen atom). 
As a consequence, $g_{11}(x,y)$ has a spherical-symmetric distribution centered at the particle position.
At a subsequent time,  $t=0.363$~fs, we start observing the formation of space corrugations, with the development of maxima and minima that guide the geodesic curve along specific regions of space.
This occurs before the trajectory reaches the scattering center and is mainly caused by the compression of the wavepacket induced by the collision with the hydrogen atom. 
The situation becomes even clearer at a later time (third panel) when the particle reaches the portion of the plane with $x<x_{H}$.
The oscillations of the metric tensor field, $g_{11}(x,y)$, are further amplified, forming a clear and well-defined path 
for the geodesic.
A similar behavior is also observed in the case of the dynamics with an impact parameter different from zero (left column of Fig.~\ref{Fig_qpot}).
The main difference is that the pattern loses the symmetry with respect to the horizontal axis passing through the scattering
 center, and the trajectory, which is confined in a different groove than in the previous case, is scattered with a different angle.
This geometrical picture of self-interference differs strongly from that associated to the Schr\"odinger interpretation, which describes the scattering
pattern as the superposition of outgoing wave fronts passing from opposite sides of the scattering center and interfering 
constructively and destructively at different angles.

\begin{figure}[h]
\begin{center}
\includegraphics[width=8.cm]{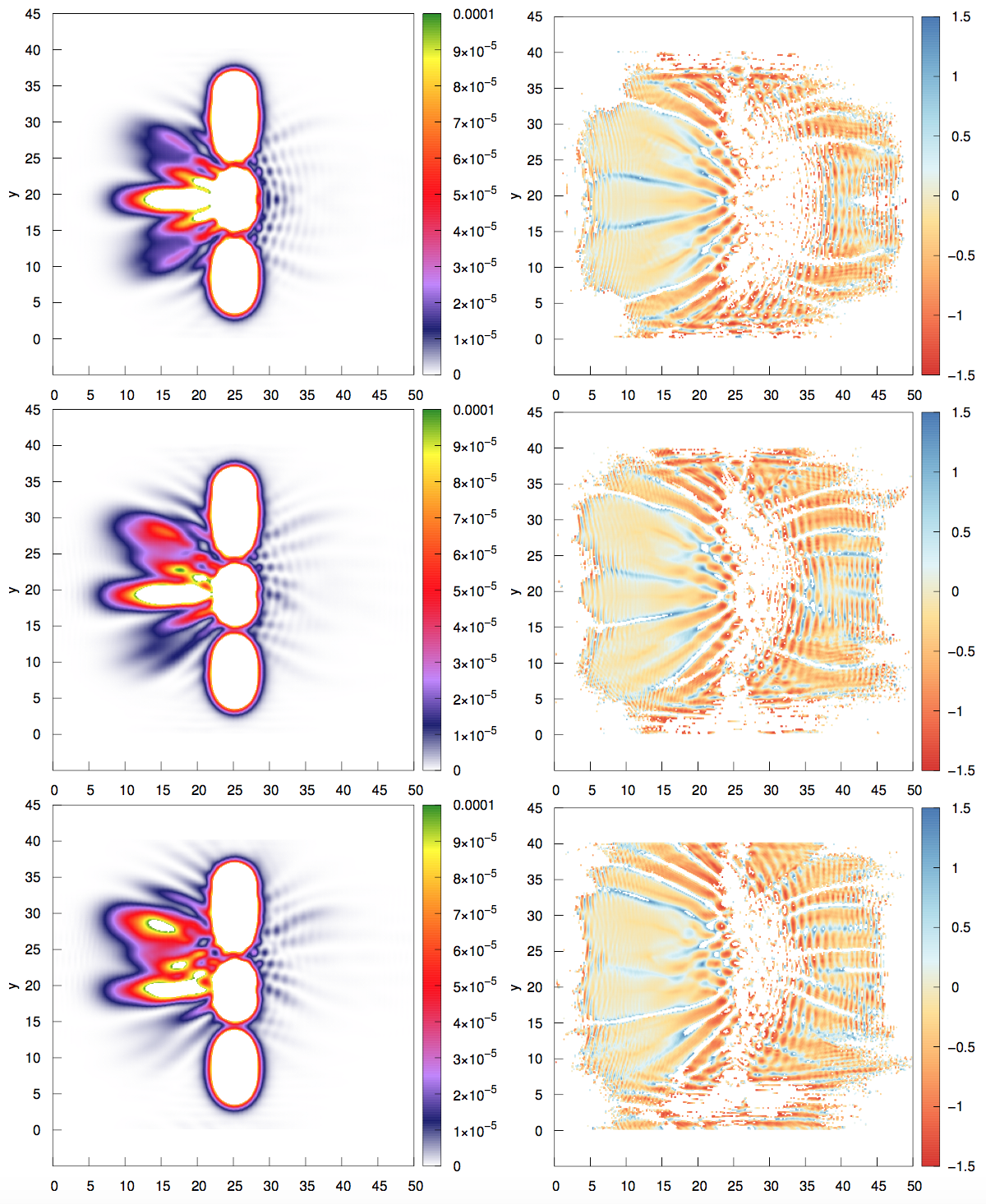}
\caption{
Left column: Two-dimensional electron density profiles of an electron passing through a double-slit
with impact parameters $b=b_0,b_e$ and $b_c$ (top to bottom).  
Right column: corresponding metric component $g_{11}(x,y)$.
All profiles are evaluated in the $xy$-plane that passes through the center of the simulation box; for more details see Fig.~\ref{figure1}. (Coordinate axes in \AA).
}
\label{Fig_slit_dens_qpot}
\end{center}
\end{figure}

\begin{figure}[h]
\begin{center}
\includegraphics[width=8.cm]{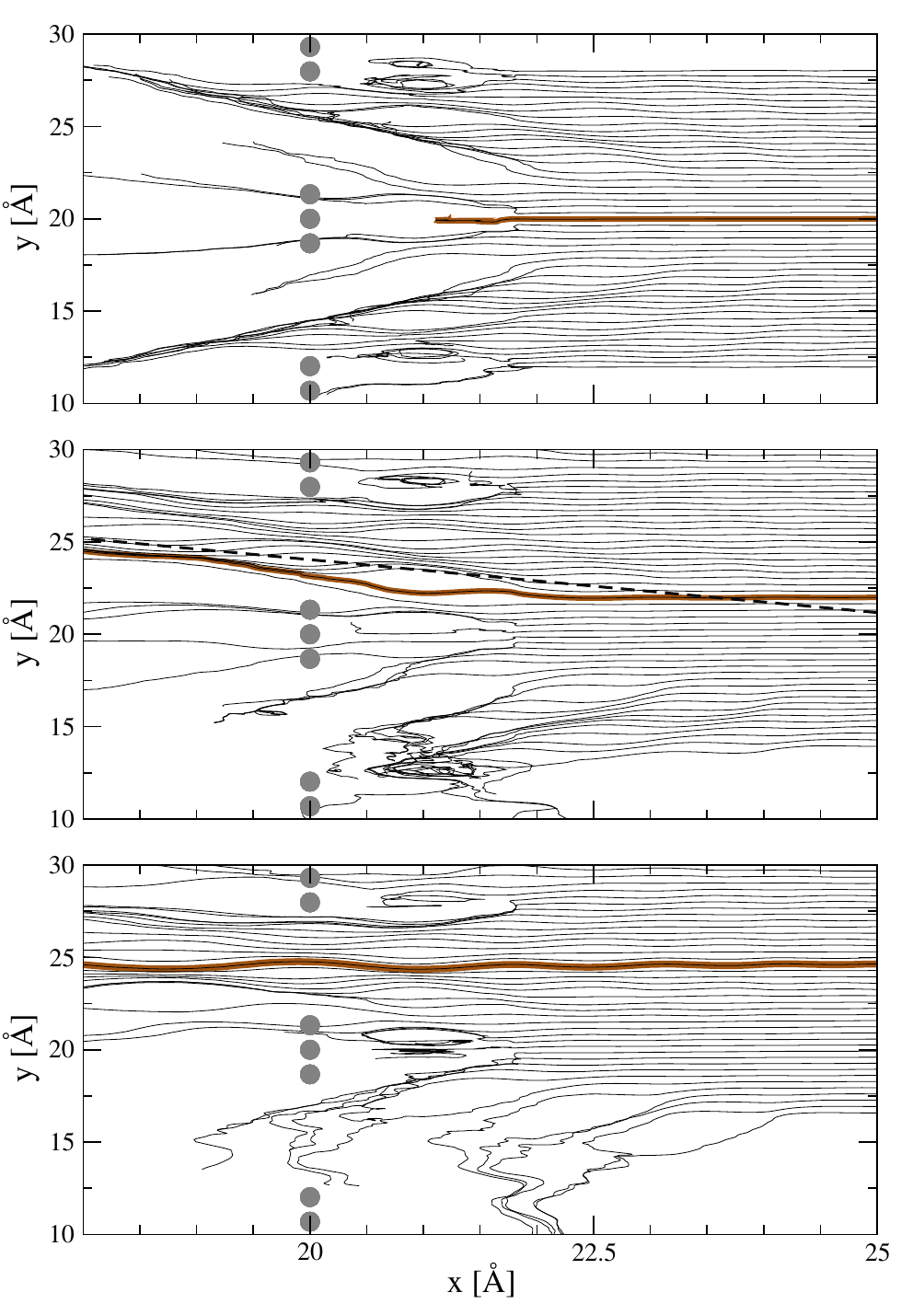}
\caption{Bohmian trajectories in the $xy$-plane obtained for the scattering of the electron by a double-slit.
The trajectories are started from a regular 1D grid passing through the center of the Gaussian at $t=0$ and 
parallel to the $y$-axis. The brown line corresponds to the geodesic. Crossings among trajectories are only apparent and are caused by the projection into the $xy$-plane.}
\label{Fig_slit_traj}
\end{center}
\end{figure}

\begin{figure}[h]
\begin{center}
\includegraphics[width=8.cm]{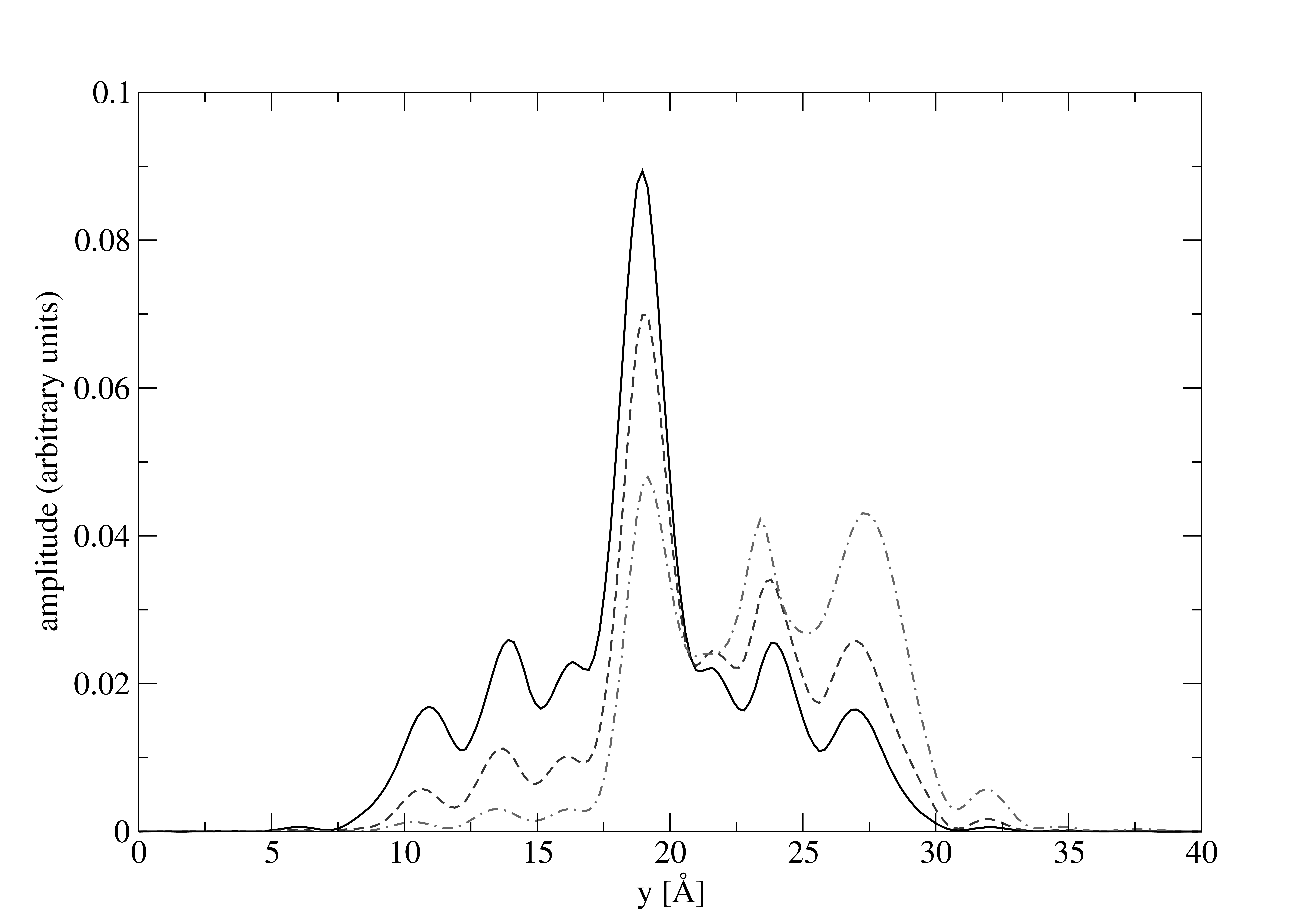}
\caption{Interference pattern on a detector screen placed at 2 \AA~ from the double-slit plane (and parallel to it)
for different values of the impact parameter $b$: $b_0$ (black line), $b_e$ (dashed line), and $b_c$ (dashed-dotted line).}
\label{Fig_ampli}
\end{center}
\end{figure}

\paragraph*{Scattering on a double-slit.}
Fig.~\ref{Fig_slit_dens_qpot} shows the self-interference pattern obtained from the scattering of an electron on a double-slit carved in a zinc monolayer surface. 
On the left-hand side, the electronic densities projected on a plane perpendicular to the surface and parallel to the 
$xy$-plane (see Fig.~\ref{figure1}) is reported for different values of the impact parameter $b$. 
From top to bottom: $b=b_0$, $b=b_e$, and $b=b_c$, where $b_0=0$, $b_e$ corresponds to the impact with the lower edge of the upper slit, and $b_c$ corresponds to the case of a wavepacket guided towards the midpoint of the upper slit. 
As expected, the electron wavefunction reproduces the typical self-interference pattern of wave-mechanics
that, according to the Copenhagen interpretation of the measurement process, 
encodes the probability distribution of the scattered electron.
The right column of Fig.~\ref{Fig_slit_dens_qpot} reports the corresponding 2D cuts in the $xy$-plane of $g_{11}(x,y)$.
Also in this case, we observe the characteristic pattern made of grooves that guide the trajectories towards the maxima of the
interference pattern.
Interestingly, the overall profile of the metric tensor $g_{11}(x,y)$  is very similar for the three values of the impact parameter, in agreement with the experimental observation of the invariance of the interference pattern from the alignment of the incoming electron with the double-slit. 
In addition, we also observe a back scattering of the space distortion that propagates in the direction opposite to that
 of the incident electron.
The corresponding Bohmian trajectories for the three values of the impact parameter $b$ are shown in Fig.~\ref{Fig_slit_traj}.
The trajectories are started from a regular grid in the $yz$-plane passing through the center of the Gaussian wavepacket at time $t=0$.
The geodesics (one for each run) corresponding to Eq.~\eqref{prop1} are highlighted in brown.
At $b=0$ (top panel), only a few trajectories continue straight, contributing to the central interference peak. 
Note that, in the Bohmian picture, these trajectories would carry the largest amplitudes, whereas those
sampled from the tail of the Gaussian distribution would be associated with lower amplitudes.
In addition, most of the trajectories contributing to the central peak originate from points situated in the front part of the wavepacket~\cite{Sanz08}.
At $b=b_e$ (middle panel), most of the trajectories are deflected at an angle of about $30^\circ$, whereas at $b=b_c$ (bottom panel), most of the transmitted trajectories continue following, roughly, a straight path.
Using wave mechanics, I compare the simulated diffraction angles with the corresponding theoretical values.
With a de Broglie wavelength $\lambda_{\text{dB}}=h/(m_e v)$ (using $v=|\bs v|=2.18 \, 10^6$ fs) the scattering angle is given by $\theta=\arcsin(\lambda/d)$, where $d$ is the separation of the virtual emitting points placed within the slits. 
Taking a value of $d$ within the two extremes $d_1=2.96$~\AA~(distance between the closest edges) and $d_2=7.94$~\AA~  (distance between the edges furthest apart) for the first scattering peak, one obtains a range of scattering angles $\theta_2=20^\circ < \theta < \theta_1=47^\circ$, which is in good agreement with the scattering angle of about $30^\circ$ measured for both the Bohmian and the geodesic trajectories (dashed line in Fig.~\ref{Fig_slit_traj}, middle panel).
Finally, I compute the electron probability distribution (diffraction pattern) on a hypothetical detector plate parallel 
to the $yz$-plane and placed 5.82~\AA~  away from the double-slit.
The density profiles for the three values of the impact parameter $b$ are shown in Fig.~\ref{Fig_ampli}.
Interestingly, the positions of the maxima are roughly independent of the value of $b$, whereas the intensity of the peaks 
shifts progressively from the center to the higher-order maxima as the impact parameter increases. 
In agreement with previous estimations of the scattering angles, the first scattering peak is under an angle of about $29^\circ$.

\section{Summary and conclusions}

The interference of single-molecule beams with a double-slit is one of the key experiments in quantum dynamics
that embodies the essence of the particle-wave interpretation of quantum mechanics. 
As stated in the Introduction, this experiment is characterized by several fundamental length scales: the de Broglie wavelength $\lambda_{\text{dB}}=h/p$, where $p$ is the modulus of the particle momentum, the dimensions of the slits (spacing $d_s$ and width $w_s$), and
the transverse coherence length, $l_c$, which correlates with the transversal spread of the molecular wavepacket and its support.
The de Broglie wavelength is a purely dynamical property of quantum systems associated with the wavepacket group velocity that
manifests itself as an interference pattern upon collision with an obstacle.
Crucial for the manifestation of the wave nature of the scattered particle is, however, the ratio between the coherence length
and the slit spacing $l_c/d_s$, which needs to be larger than $1$ to guarantee the passage of the wavepacket 
through at least one pair of slits (the minimum requirement for the formation of an interference pattern according to wave-mechanics).
This condition implies the preparation of coherent single-particle beams with values of $l_c$ that are several orders of magnitude larger than the
`classical' dimension of the scattered particle. 
For example, in the case of the experiment with buckyballs~\cite{Zeilinger1999,Nairz2003}, $d_s$ is on the order of $55$ nm, whereas the `classical' diameter of C60 is about $0.71$ nm. 
Experimentally, this condition is reached through a cooling scheme that allows one to trap and confine the particle's wavefunction so tightly in space that, once the particle is released, its large momentum uncertainty ensures a fast expansion in the transverse direction.
To keep a transverse coherence over such long distances, the experiments need to be carried out in complete isolation to avoid interferences and decoherence effects. 

Several interpretative schemes have been developed to account for this fundamental experiment in QD: the wave picture and its
reformulations in terms of path integrals, the deterministic de Broglie--Bohm trajectories, and the geodesic formulation in curved space. 
In the following, I will briefly comment on the implications these descriptions of QD have for the interpretation of the self-interference experiment.

\paragraph{The wave picture of QD.}
Quantum mechanics was originally formulated in terms of amplitude fields, i.e., the system wavefunctions that 
give rise to a `probabilistic' interpretation of the physical processes and measurements.
To reconcile this picture with the emergence of the `classical' deterministic world,
an interpretative mechanism was introduced that describes a measurement as an event that causes the reduction of 
the QM probability density to a single, real valued, observable through a wavefunction collapse 
(e.g., the particle-wave duality of the Copenhagen interpretation). 
Several attempts have been made to  mathematically justify the `information collapse' associated with
the classical measurement process, but their effectiveness is still being debated~\cite{Hepp72}. 
The wavefunction collapse  therefore remains a fundamental open problem in the theoretical foundations of QM, as also stated by J.S. Bell:
``so long as the wave packet reduction is an essential component, 
and so long as we do not know exactly when and how it takes over from the Schr\"odinger equation, we do not have an exact and unambiguous understanding of our most fundamental physical theory"~\cite{Bell88}. 


\paragraph{Feynman's path integrals.}
The most commonly accepted trajectory-based interpretation of non-relativistic quantum mechanics is 
probably the one based on the  path integral formalism introduced by R. Feynman.
Within this framework, the transition from an initial point in the configuration space ($q_i,t_i$) to a final
one ($q_f,t_f$) is described by a quantum-mechanical amplitude $K(f,i)$ that is given by the coherent 
sum of all possible \textit{independent} paths connecting the two extrema; formally
\begin{equation}
K(f,i)= \sum_{\mathcal{P}[q(t)]} \mathcal{N} \, e^{(i/\hbar) S_C[q(t)]} \, ,
\label{Feynman_prop}
\end{equation}
where $\mathcal{P}[q(t)]$ is the set of all paths connecting \textit{i} to \textit{f}, 
 $S_C[q(t)]$ is the classical action $S_C(q_0,t_0,q_f,t_f)=\int_{t_0}^{t_f} 
\left[ \frac{m}{2} (\dot{q}(t))^2 - V(q(t)) \right] \, dt$, and $\mathcal{N}$ a normalization factor~\cite{Feynman_Hibbs}.
The link to the wavefunction (amplitude field) description of conventional quantum mechanics is
then established by relating the amplitude $K(f,i)$ to the probability $P(f,i)$ for the transition from $i$ to $f$,
$P(f,t)=|K(f,i)|^2$, which leads to the identification of $K(f,i)$ with the QM propagator (for any initial time $t_i$)
\begin{equation*}
\Psi(q_f,t_f) = \int d q_i  \, K(q_f, t_f;q_i,t_i) \Psi(q_i,t_i) \, .
\label{Feynman_wave_eq}
\end{equation*}
It is interesting that Feynman's interpretation of the paths goes beyond the formulation
of the QM propagator; as stated in his classical book on QED, Feynman was also a strong supporter of the 
particle interpretation of fundamental physics: ``I want to emphasize that light comes in this form -- particles. It
is very important to know that light behaves like particles, especially for those you have gone to school,
where you were probably told something about light behaving like waves. I'm telling you that the way 
it \textit{does} behave -- like particles''.
Despite this, it is hard to identify the role of the physical paths within the path integral formalism, since no precise prescription is provided on how these paths are constructed 
from the initial conditions. 
Only after the collection of all possible paths summed according
to Eq.~\eqref{Feynman_prop} does a probabilistic (non-deterministic) picture of the `particle' dynamics emerge.
Through its connection with the particle-wave description, the path integrals formulation of QD inevitably inherits all 
issues related to the interpretation of the double-slit experiment in terms of wavefunctions, in particular the requirement of a
transverse coherence over spatial distances that largely exceed the `classical' size of the system.

\paragraph{De Broglie--Bohm's trajectories.}
The de Broglie--Bohm description of quantum mechanics is half way between a deterministic trajectory-based
approach and a wavefunction-based interpretation. 
More precisely, within this formalism, `quantum' trajectories associated with point particles are guided by a 
quantum-mechanical wave that, in the non-relativistic limit, evolves according the time-dependent Schr\"odinger equation.
The application of Bohmian dynamics in the description of the quantum interference in Young's two-slit experiment
is discussed in~\cite{Philippidis_1979,Philippidis_1982} and more recently in~\cite{Sanz15}. 
These studies provide a detailed analysis of the role of the quantum potential and phase in explaining
self-interference and the related Wheeler's delayed choice paradox~\cite{Wheeler78,Sanz15}.
What makes Bohmian dynamics deterministic is the introduction of a set of so-called \textit{hidden variables} that 
complement the quantum-mechanical description of reality, rendering its probabilistic interpretation unnecessary.
The literature on the meaning and interpretation of hidden variables is very rich and there is no need here to
address this subject. 
Interestingly, despite several attempts (see von Neumann~\cite{Neumann_book}), there still are no solid 
theoretical objections about the possibility of complementing `standard' QM with additional hidden variables 
that can render the theory deterministic. 
Despite the introduction of the pilot wave guiding the trajectory, the de Broglie--Bohm dynamics remains essentially
a deterministic theory and \textit{per se} has no probabilistic content.
However, as in the case of Feynman's path integrals, it is possible to connect the de Broglie--Bohm theory with 
conventional wavefunction-based quantum dynamics.
By sampling a number of initial configuration space points distributed according to a given wavepacket 
at an initial time $t=0$, the time evolution of this ensemble (according to Eq.~\eqref{eq:vel_field}) exactly reproduces the distribution 
of the same wavepacket at any later time $t>0$, as if it had propagated by the Schr\"odinger's time-dependent 
equation. Recovering a probabilistic description from its deterministic formulation is always much simpler than 
the inverse process~\cite{Bohmian_note}.

\paragraph{Geometrical description of QD: Geodesic paths.} 
In this picture, the quantum dynamics of point particles is formulated in terms of deterministic trajectories 
evolving in a curved Finsler space, whose curvature is determined by a \text{nonlocal} matter-energy tensor.
The fundamental characteristics of this theory can be summarized as follows:
(\textit{i}) The wave nature of QD is absorbed into the geometrical properties of the Finsler space, providing a way
to resolve the \textit{particle-wave} duality in QM~\cite{Bhom_vs_geom}.
While the de Broglie-Bohm dynamics is formulated in a flat space where both point particle and guiding wave (field) propagate with their equation of motion, in the geometric approach the physical object is a point particle that moves along a geodesic in a curved space. 
Also in this case, there is a \textit{field} associated to the particle that defines the Finsler metric~\cite{tavernelli_geom}. 
However, differently from de Broglie-Bohm dynamics, this field describes a geometric tensor and is not associated to a physical quantity, like the pilot wave~\cite{note_on_quantum_gravity}.
(\textit{ii}) The appearance of a nonlocal metric tensor that depends on all particle coordinates (position and internal degrees of freedom, such as spin, which can be associated with the hidden variables) agrees 
with Bell's theorem~\cite{Bell88}, which states that QM cannot be completed by a local hidden-variable theory because it is
intrinsically nonlocal and does not satisfy the principle of local realism~\cite{Bell88,Bohm_Hiley}.
(\textit{iii}) As in Bohmian mechanics, the geometrical formulation reproduces the predictions and measurements of quantum mechanics and, in particular, the electron self-interference pattern studied in this work.
(\textit{iv}) The separation of the particle (trajectory) from the wave (space curvature field) character of matters allows one to bypass 
the concept of wavefunction collapse introduced in the Copenhagen interpretation of QM. 
(\textit{v}) Concerning self-interference, the geometric interpretation of QD does not require the assumption of gigantic transverse 
coherence lengths ($l_c$) as it is necessary for the conventional wavefunction-based interpretation of double-slit experiments.
As shown in the case of the electron scattering at a double-slit, self-interference arises from the Finsler space curvature induced by the action of the nonlocal quantum potential. The projection of the metric tensor on the plane perpendicular to the slit (see Fig.~\ref{Fig_slit_dens_qpot}) shows well-defined paths that guide the trajectories to form the characteristic interference pattern.

In conclusion, we showed that the geometrical interpretation of QD offers a valid description of the self-interference
process in scattered electrons.
This approach introduces determinism in the characterization of the particle dynamics while incorporating quantum 
nonlocality and wave behavior into the \textit{immaterial} geometry of the space.
The identification of quantum effects with the purely geometrical properties of space and time opens up new avenues for a deterministic interpretation of QM and its unification with gravitation~\cite{covariance,tavernelli_quant_grav}.

\begin{acknowledgments}
The author acknowledges  the Swiss SNF grant 200020-130082  for funding and the Center for Advance Modeling Science (CADMOS) for computer time at the IBM BlueGene/Q. The author also thanks Angel S. Sanz for interesting discussions on
the interpretation of self-interference within  Bohmian dynamics.
\end{acknowledgments}

\appendix

\renewcommand{\theequation}{A\thesection.\arabic{equation}}
\setcounter{equation}{0}
\section{Appendix A}

The Finsler metric in Eq.~\eqref{prop2} defines a dynamical system through the minimisation of the `action' functional $I(\gamma)=\int_{\tau_1}^{\tau_2} \Lambda(x, y)\, d\tau$ for a path $\gamma$ and given initial and final conditions ($\gamma(\tau_1)$ and $\gamma(\tau_2)$), when the  following three conditions are fulfilled~\cite{Caratheodory,tavernelli_geom} 
(\textit{i}) positive homogeneity of degree one in the second argument, $\Lambda(x,k y)= k \Lambda(x, y), k>0 $, 
(\textit{ii}) $\Lambda(x,y) > 0$ with $\sum_i (y^i)^2 \neq0$, and 
(\textit{iii}) $\frac{1}{2} \frac{\partial^2 \Lambda^2(x,y)}{\partial y^{\alpha} \partial y^{\beta}} \xi^{\alpha} \xi^{\beta} > 0, \forall {\xi}\neq \lambda y$. 
As a reminder, I use $x=(x^0, \bs q), y=(y^0, \dot{\bs{q}}), x^0=t, y^0=\partial t / \partial \tau$, while $\tau$ measures the progression of the dynamics. 
\paragraph*{Condition (i)} is fulfilled by construction. 
\paragraph*{Condition (ii)} is less restrictive than it appears. As discussed in Ref.~\cite{Rund59}, it is in fact always possible to add to the integrand of the action functional an arbitrary function $S(x)$ of class $C^1$, which makes the integrand positive in the region of space of interest, $\Omega$. 
The quantity $(\partial S(x)/\partial x^i) \, dx^i$ is then an exact differential and therefore the integral along any curve $\gamma$ joining the initial and final points, $x_1$ and $x_2$, gives a constant value $S(x_2)-S(x_1)$, independently from the path (with the condition that it lies in $\Omega$). 
As a consequence, if I add the term $(\partial S(x)/\partial x^i) y^i$ to the integrand $\Lambda(x, y)$,
the extremal path is identical to the one obtained using the original action. 
\paragraph*{Condition (iii)} is more difficult to prove in general. 
In the case of one-particle systems, it is possible to derive an equivalent inequality that can be easily verified. 
First, I include the external potential $V(q)$ (where $x=(x^0,q)$) as part of the total potential that determines the space geometry. 
This is always possible, especially when all interaction potentials are quantized (as quantum fields) and combined into a generalized quantum potential~\cite{hollandbook}.
I will call this quantity, the generalized quantum potential $Q'(x)$.
In the one dimensional case (but the argument can be extended to the 3D case since $\partial_{\dot{{q}}} ((m |\dot{{q}}|^2/2) = m \, \dot{q}$ in any dimension), condition (\textit{iii}) can be reformulated as
\begin{equation}
\det
\begin{pmatrix}
\partial_{ij} \Lambda & \partial_i  \Lambda \\
\partial_i \Lambda & 0 
\end{pmatrix}
=
\begin{vmatrix}
\partial_{ij} \Lambda & \partial_i  \Lambda \\
\partial_i \Lambda & 0 
\end{vmatrix}
<0
\end{equation}
where $i \in \{ 0,1\}$ ($\partial_i=\partial_{\dot{x}^i}$, $\partial_{ij}=\partial_{\dot{x}^i} \partial_{\dot{x}^j}$). 
For $\Lambda( x, y)$ defined as
\begin{equation}
\Lambda(x, y) = \mathcal{T}(\dot q) /y^0 - Q'( q) y^0
\end{equation}
(remind that $x\equiv(x^0,q^1), y \equiv (y^0, \dot q^1)$, with $y^0=\dot x^0=\partial t/\partial\tau$) one gets
\begin{equation}
\begin{vmatrix}
 1 / (y^0)^3 \mathcal{T}          & - m \dot q^1 /(y^0)^2    &  - \mathcal{T}/ (y^0)^2 - Q'\\
- m \dot q^1 /(y^0)^2              & m /y^0                           &  m \dot q^1 /y^0 \\
- \mathcal{T}/ (y^0)^2 - Q' & m \dot q^1 /y^0             & 0
\end{vmatrix}
<0
\end{equation}
where $\mathcal{T}=(1/2) m (\dot q^1)^2$.
This implies (for $m>0$ and $y^0>0$)
\begin{equation}
Q'< \frac{m (\dot q^1)^2 (1-\sqrt{2})}{2 (y^0)^2} \quad 
\end{equation}
and therefore
\begin{equation}
\frac{m (\dot q^1)^2}{2 (y^0)^2} + Q' < \frac{m (\dot q^1)^2 (2-\sqrt{2})}{2 (y^0)^2} \, .  
\end{equation}
Note that $y^0=1$ in Euclidean space (i.e., in classical dynamics), and in all quantum simulations done so far this condition is violated only up to a few percents.
Since $\frac{(\dot q^1)^2 (2-\sqrt{2})}{2 (y^0)^2}>0$, the requirement for the total energy of the system
\begin{equation}
\frac{ m (\dot q^1)^2}{2} + Q' < 0 
\end{equation}
(assuming $y^0 \sim 1$) directly validates condition (\textit{iii}).

\renewcommand{\theequation}{B\thesection.\arabic{equation}}
\setcounter{equation}{0}
\section{Appendix B}

In this appendix, I review the main aspects of Finsler geometry that are related to the subject of this work.
For a more detailed account on this subject the interested reader should refer to the specialized literature~\cite{Rund59,Thesis_Pfeifer}.

A point in the tangent bundle TM is represented by the coordinates $(x^0,\dots,x^{{n}}, y^0, \dots, y^{{n}})$, where $M$ is the base manifold of dimension $n+1$. The tangent space of TM in a point $x=u$ is described by the coordinated $(\frac{\partial}{\partial x^0}=\partial_0, \dots, \frac{\partial}{\partial x^{n}}=\partial_{n}, \frac{\partial}{\partial y^1}=\bar\partial_1, \dots, \frac{\partial}{\partial y^{n}}=\bar \partial_{n})$.

For a Finsler space ($M$,$F$) defined by the configuration space manifold $M$ and the Finsler's function $F$ the (0,2)-d metric tensor field is defined as
\begin{equation}
g_{ab}(x,y)=\frac{1}{2} \bar\partial_a \bar\partial_b F^2(x,y) \, ,
\end{equation}
while the (0,3)-d Cartan tensor is given by the third derivative of $F^2$ with respect to the tangent space coordinates $y$
\begin{equation}
C_{abc}(x,y)=\frac{1}{4} \bar\partial_a \bar\partial_b \bar\partial_c F^2(x,y) \, .
\end{equation}
(In case $C_{abc}(x,y)=0$ everywhere in the tangent space,  the Finsler space becomes a metric space with $g_{ab}(x)$ independent on tangent space  coordinates $y$).
The corresponding non-linear Cartan connection is defined by the coefficients 
\begin{equation}
?N^{a}_{}b?(x,y)=?\Gamma^a_{}{bc}?(x,y) y^c - ?C^a_{}{bc}?(x,y) ?\Gamma^c_{}{pq}?(x,y) y^p y^q
\label{Nab_def}
\end{equation}
(with $?C^a_{}{bc}?(x,y)=g^{ad}(x,y)?C_{}{dbc}?(x,y)$) or, in a more compact form,
\begin{equation}
?N^{a}_{}b?(x,y)=\frac{1}{2}\bar\partial_b (?\Gamma^a_{}{cd}?(x,y) y^c y^d) 
\label{non-lin-conn}
\end{equation}
where $g^{ab}(x,y)$ is the inverse of $g_{ab}(x,y)$ and $?\Gamma^a_{}{bc}?= g^{aq}(\partial_b g_{qc} + \partial_c g_{qb} - \partial_q g_{bc}) $ (removing the dependence on $x$ and $y$).
The non-linear curvature derived from $?N^{a}_{}b?$ is 
\begin{equation}
?R^a_{}bc?=\delta_c ?N^{a}_{}b? - \delta_b ?N^{a}_{}c? \, .
\end{equation}
The connection allows us to decompose the tangent space $T_u$TM ($T_u\tilde{\text M}$) into two subspaces: the vertical space (V$_u$TM) tangent to T$_u$M (span by $\bar\partial_a$) and the horizontal subspace (H$_u$TM) tangent to the base space manifold M as defined by the connection $?N^{a}_{}b?$.
In coordinates, this definition implies the transformation $\{\partial_a, \bar\partial_b\} \rightarrow \{\delta_a=\partial_a -?N^{b}_{}a?\partial_b, \bar\partial_b \}$. 
Regarding the tangent bundle as a manifold in its own $\tilde{M}=TM$, one can associate linear covariant derivatives to this manifold, which are compatible with the structured induced by the non-linear connection and that preserve the horizontal-vertical split of its tangent bundle $T\tilde{M}$ with basis $\{\delta_a,\bar\partial_b\}$, causing no mixing. 
The linear covariant derivative  in the horizontal-vertical basis is
\begin{align}
\tilde\nabla_{\delta_a} {\delta_b} &= ?{{\tilde\Gamma}}^c_{}{ab}? \delta_c \\
\tilde\nabla_{\delta_a} {\delta_{b'}} &= ?{{\tilde\Gamma}}^{c'}_{}{ab'}? \bar{\partial}_{c} \\
\tilde\nabla_{\delta_{a'}} {\delta_b} &= ?{{\tilde Z}}^c_{}{a'b}? \delta_c \\
\tilde\nabla_{\delta_{a'}} {\delta_{b'}} &= ?{{\tilde Z}}^{c'}_{}{a'b'}? \bar{\partial}_{c} 
\end{align}
where $a,b,c=0,\dots, n$; $a',b'=n+1,\dots,2 (n+1)$;  $n+1$ is the dimension of M and 
\begin{align}
?{{\tilde \Gamma}}^c_{}{ab}? &= \frac{1}{2} g^{cq} (\delta_a g_{bq} + \delta_b g_{aq} - \delta_q g_{ab})\\
?{{\tilde Z}}^c_{}{ab}? &= g^{cq} C_{abq} \, .
\end{align}
Note that while the Cartan non-linear connection is unique, linear connections are not and therefore alternative
definitions are also possible~\cite{Thesis_Pfeifer}.
The horizontal part of the curvature $^l\hspace{-0.07cm}R(\delta_a, \delta_b)(.)$  is then given by
\begin{align}
^l\hspace{-0.07cm}?R^q_{}{cab}?=\delta_a ?{{\tilde\Gamma}}^q_{}{cb}? - 
\delta_b ?{{\tilde\Gamma}}^q_{}{ca}? + 
?{{\tilde\Gamma}}^q_{}{ma}? ?{{\tilde\Gamma}}^m_{}{cb}? - 
?{{\tilde\Gamma}}^q_{}{mb}?  ?{{\tilde\Gamma}}^m_{}{ca}? 
- ?C^q_{}{cm}? ?R^m_{}{ab}? \, ,
\end{align}
while the relation to the non-linear curvature becomes
\begin{equation}
?R^q_{}{ab}?=- \, ^l\hspace{-0.07cm}?R^q_{}{cab}? y^c \, .
\end{equation}
The corresponding geodesic equation for a curve $\tau \mapsto \gamma(\tau)$ is
\begin{equation}
\ddot{\gamma}^a + ?N^{a}_{}b?(\gamma,\dot{\gamma}) \dot\gamma^b= 0 \, ,
\end{equation}
which is equivalent to Eq.~\ref{prop1} (for the case of zero external potential). 
The proof requires the definition of a new tensor, $?P^a_{}bc?$, defined as
\begin{equation}
?P^a_{}bc? \, y^c=?N^{a}_{}b?.
\end{equation}
According to Eq.~\ref{Nab_def}, this tensor satisfies the equation 
\begin{equation}
?P^a_{}bc?= ?\Gamma^a_{}{bc}? - ?C^a_{}{bd}? ?\Gamma^d_{}{qc}? y^q \, ,
\end{equation}
from which (using the identities $?C_{}{abc}? \, y^a = ?C_{}{abc}? \, y^b = ?C_{}{abc}? \, y^c=0$) one obtains
\begin{equation}
?P^a_{}bc? \, y^k= ?\Gamma^a_{}{bc}? \, y^k \, .
\end{equation}
Therefore, from the geodesic equation
\begin{equation}
\ddot{\gamma}^a + ?\Gamma^{a}_{}bc?(\gamma,\dot{\gamma}) \dot\gamma^b \dot\gamma^c= 0
\end{equation}
one first gets
\begin{equation}
\ddot{\gamma}^a + ?P^{a}_{}bc?(\gamma,\dot{\gamma}) \dot\gamma^b \dot\gamma^c= 0
\end{equation}
and finally
\begin{equation}
\ddot{\gamma}^a + ?N^a_{}b?(\gamma,\dot{\gamma}) \dot\gamma^b = 0 \, .
\end{equation}


\begin{thebibliography}{47}
\bibitem[{not()}]{note_cayley}
\bibinfo{note}{For a small time-step $\delta t$, \begin{equation} \phi_k(\bs
  q,t+\delta t)=\exp(-i \delta t \, \hat{H}_{0}[\left\{ \phi_{i} \right\}])
  \phi_k(\bs q,t) + \mathcal{O}(\delta t^2) \, , \label{eq:numerics1}
  \end{equation} where the propagation operator $\exp(-i \delta t \,
  \hat{H}_{0}[\left\{ \phi_{i} \right\}])$ is replaced by its finite difference
  representation \begin{equation} \exp(-i \delta t \, \hat{H}_{0}[\left\{
  \phi_{i} \right\}])= \frac{1-(i \delta t/2) \hat{H}_{0}[\left\{ \phi_{i}
  \right\}]}{1+(i \delta t/2) \hat{H}_{0}[\left\{ \phi_{i} \right\}]} +
  \mathcal{O}(\delta t^2) \end{equation}}.
\bibitem[{Boh()}]{Bohmian_note}
\bibinfo{note}{De Broglie--Bohm trajectories differ substantially from Feynman
  path integrals. In particular, the trajectories in the two-slit interference
  experiment follow remarkably different paths, even though their distributions
  (obtained through the weighted sum in Eq.~\eqref{Feynman_prop} for the path
  integrals) reproduce the correct intensity patterns in both cases.
  Physically, while there is an ensemble of possible Feynman paths connecting
  an initial point in the configuration space ($r_0,t_0$) to a final one
  ($r_f,t_f$) (which needs to be added coherently), only a single one
  corresponds to the Bohmian path (the one that minimizes the Bohmian action
  $S_B(r_0,t_0,r_f,t_f)=\int_{t_0}^{t_f} \left[ \frac{1}{2m} (\dot{r}(t))^2 -
  V(r(t)) - Q(r(t))\right] \, dt$. This means that for any single-valued
  pilot-wave Bohmian, paths starting from different initial conditions
  ($r_0,t_0$) cannot cross in the configuration space.}
\bibitem[{Bho()}]{Bhom_vs_geom}
\bibinfo{note}{\textcolor{blue}{ The main differences between the Bohmian and
  geometrical formulations of quantum dynamics can be summarized as follows:
  (i) the nature of the ``driving force'': in Bohmian dynamics the trajectories
  are guided by a wavefunction of unspecified nature, while in the geometrical
  approach it is associated to a physical curvature of space. (ii) the
  statistical interpretation of the initial conditions: in Bohmian dynamics the
  initial particles distribution is associated to the spread of the system
  wavefunction. In this picture, different wavepackets that share part of their
  support in position space can contribute to a trajectory with the same
  deterministic initial conditions. In the geometrical interpretation, each
  particle is associated to an unique point in space, while the wave-nature of
  the dynamics is confined to the propagation of the space curvature}}.
\end{thebibliography}
\begin{thebibliography}{51}
\bibitem[{not({\natexlab{a}})}]{note_cayley}
\bibinfo{note}{For a small time-step $\delta t$, $ \phi_k(\bs
  q,t+\delta t)=\exp(-i \delta t \, \hat{H}_{0}[\left\{ \phi_{i} \right\}])
  \phi_k(\bs q,t) + \mathcal{O}(\delta t^2) \, , $
  where the propagation operator $\exp(-i \delta t \,
  \hat{H}_{0}[\left\{ \phi_{i} \right\}])$ is replaced by its finite difference
  representation $ \exp(-i \delta t \, \hat{H}_{0}[\left\{
  \phi_{i} \right\}])= \frac{1-(i \delta t/2) \hat{H}_{0}[\left\{ \phi_{i}
  \right\}]}{1+(i \delta t/2) \hat{H}_{0}[\left\{ \phi_{i} \right\}]} +
  \mathcal{O}(\delta t^2) $}.
\bibitem[{Bho()}]{Bhom_vs_geom}
\bibinfo{note}{The main differences between the Bohmian and
  geometrical formulations of quantum dynamics can be summarized as follows:
  (i) the nature of the ``driving force'': in Bohmian dynamics the trajectories
  are guided by a wavefunction of unspecified nature, while in the geometrical
  approach it is associated to a physical curvature of space. (ii) the
  statistical interpretation of the initial conditions: in Bohmian dynamics the
  initial particles distribution is associated to the spread of the system
  wavefunction. In this picture, different wavepackets that share part of their
  support in position space can contribute to a trajectory with the same
  deterministic initial conditions. In the geometrical interpretation, each
  particle is associated to an unique point in space, while the wave-nature of
  the dynamics is confined to the propagation of the space curvature}.
\end{thebibliography}

\end{document}